\documentclass[
    twocolumn,
    aps,
    prl,
    superscriptaddress,]{revtex4-2}
\usepackage{amsmath}
\usepackage{graphicx}
\usepackage{MnSymbol}
\usepackage{hyperref}
\usepackage{fancyhdr}
\usepackage{booktabs}
\usepackage{multirow}
\usepackage{parskip}
\usepackage{natbib}
\usepackage{enumerate}
\usepackage{float} 

\begin{document}

\title{Leakage at interfaces: a comprehensive study based on Persson contact mechanics theory}

\author{R. Xu}
\affiliation{Peter Gr\"unberg Institute (PGI-1), Forschungszentrum J\"ulich, 52425, J\"ulich, Germany}
\affiliation{State Key Laboratory of Solid Lubrication, Lanzhou Institute of Chemical Physics, Chinese Academy of Sciences, 730000 Lanzhou, China}
\affiliation{MultiscaleConsulting, Wolfshovener str. 2, 52428 J\"ulich, Germany}

\author{L. Gil}
\affiliation{BD Medical-Pharmaceutical Systems, 11 Rue Aristide-Berg{\`e}s, Le Pont de Claix 38801, France}

\author{J. Singer}
\affiliation{BD Medical-Pharmaceutical Systems, 11 Rue Aristide-Berg{\`e}s, Le Pont de Claix 38801, France}

\author{L. Gontard}
\affiliation{BD Medical-Pharmaceutical Systems, 11 Rue Aristide-Berg{\`e}s, Le Pont de Claix 38801, France}

\author{W. Leverd}
\affiliation{BD Medical-Pharmaceutical Systems, 11 Rue Aristide-Berg{\`e}s, Le Pont de Claix 38801, France}

\author{B.N.J. Persson}
\affiliation{Peter Gr\"unberg Institute (PGI-1), Forschungszentrum J\"ulich, 52425, J\"ulich, Germany}
\affiliation{State Key Laboratory of Solid Lubrication, Lanzhou Institute of Chemical Physics, Chinese Academy of Sciences, 730000 Lanzhou, China}
\affiliation{MultiscaleConsulting, Wolfshovener str. 2, 52428 J\"ulich, Germany}

\begin{abstract}

We present a comprehensive study of gas leakage at interfaces based on Persson contact mechanics theory. A prototype syringe system consisting of a rubber stopper and a glass barrel is selected, where surface roughness is characterized using measurements from stylus profilometry and atomic force microscopy, and contact pressure distributions are obtained from Finite Element Method (FEM) calculations. Leakage prediction is performed using Multiscale Contact Mechanics (MCM) software. The predicted results show good agreement with experimental measurements under controlled dry conditions. Sensitivity analyses indicate that small variations in elastic modulus and contact pressure can significantly influence leakage, particularly near the percolation threshold. This work provides a generalized and validated framework for leakage prediction and offers practical guidance for the design of sealing systems in pharmaceutical and engineering applications.

\end{abstract}

\maketitle

\thispagestyle{fancy}

%%%%%%%%%%%%%% main text %%%%%%%%%%%%%%%%

\section{Introduction}

All solid surfaces exhibit roughness. When two nominally flat surfaces come into contact, if the normal pressure is low or the elasticity of the solid is high, non-contact areas will always exist at the interface, leading to interfacial separation (gaps) between the solids. When the non-contact areas percolate, the gaps extend from one side of the nominal contact zone to the other, resulting in leak paths that allow fluid to flow through the interface.

A seal is a device used to close these gaps, prevent the leakage of fluids or gases, and ensure the proper functioning of mechanical systems. Seals are typically made of elastomeric materials (such as rubber) that can adapt to various shapes and pressure conditions. Common types include O-rings and gaskets. Seal failure can result in environmental pollution, energy loss, and significant engineering failures.
%such as issues with the Boeing Starliner or, in the worst case, the Challenger disaster, which resulted in the loss of human life.

Although sealing is of great importance, predicting leaks theoretically remains a complex challenge. The leak path depends on the interfacial separation, which in turn depends on surface roughness. Solid surfaces exhibit roughness over many decades of length scales \cite{persson2014fractal, nayak1971random, aghababaei2022roughness}, with common engineering surfaces displaying roughness spanning lengths from centimeters to nanometers. This multi-scale roughness complicates investigations using deterministic simulation methods.

One approach to addressing the problem of interfacial surface separation under a given pressure is to use contact mechanics theories. Two main families of theories are used to study the contact of rough solids: the Greenwood-Williamson (GW) theory \cite{greenwood1966contact} and the Persson theory \cite{persson2001theory, persson2002adhesion, Persson2006contact, persson2007relation, Persson2008onthe, yang2008contact}.

The GW theory approximates the roughness of a real surface as a series of spherical caps with random heights. The contact of each asperity is determined using Hertz theory, and from this, the total number of contact regions, total contact area, and surface separation at a given load can be derived. The GW theory is widely used, particularly by engineers, due to its simplicity. However, it rely on several assumptions: independent spherical asperities with identical radii and a random distribution of heights. These assumptions neglect (i) the multiscale nature of real surface roughness and, most importantly, (ii) the long-range elastic deformation. When one asperity deforms, neighboring asperities around it also deform. This effect is critical in calculating leakage paths \cite{almqvist2011interfacial}. Although refinements to the GW model, such as the Bush, Gibson, and Thomas (BGT) model \cite{bush1975elastic} and the Chang, Etsion, and Bogy (CEB) model \cite{chang1987elastic}, have been developed to address these limitations, these theories are generally considered suitable only for small loads \cite{carbone2011contact}. Due to these inherent limitations, particularly the neglect of elastic coupling and multiscale roughness, the GW model has been proven not suitable for leakage analysis, as it cannot reliably predict surface separation and the size of leak paths \cite{almqvist2011interfacial,dapp2012self}.

The Persson theory discards the assumption of a small real contact area relative to the nominal contact area and instead begins with the case of complete contact \cite{persson2001theory, yang2008contact}. The central idea is that at low magnification $\zeta$, the true contact area $A$ equals the nominal contact area $A_0$. Here, $A$ and $A_0$ refer to the projections of the actual and nominal contact regions on the horizontal $xy$ plane. However, as the magnification increases, non-contact areas (due to surface roughness) become visible. The Persson theory can be used to calculate various contact properties, such as the relative contact area $A/A_0$ and the probability distribution of interfacial separation $P(u)$, which is used in the theory for leakage.

While the Persson theory is exact under full contact conditions \cite{carbone2011contact}, it provides accurate estimations even at small contact pressures \cite{yang2008contact, almqvist2011interfacial}, making accurate leakage predictions in contacts possible \cite{persson2008theory, lorenz2010dependence, dapp2012self}.

Leakage prediction using the Persson theory requires the surface roughness power spectral density (calculated from measured surface topography) and information about the contact as input. This includes the elastic (or elastoplastic) properties of the solids, viscosity properties of the fluid, contact pressure, and fluid pressure (or pressure difference). In recent years, numerous practical studies have applied the Persson contact mechanics theory to investigate leakage and sealing. These include research on static metallic seals \cite{persson2016leakagemetal2, fischer2020metal1, fischer2021influence}, elastomer seals \cite{tiwari2017rubber, akulichev2018interfacial, persson2022fluid, rodriguez2022air, huon2022air}, metallic ball valves \cite{fischer2023ballvalve}, dynamic seals \cite{bauer2022pneumaticvalve, peng2021investigation, wang2024numerical}, and leakage in face masks \cite{persson2021mask}.

We summarize the state of the art in using the Persson contact mechanics theory to study leakage and sealing problems as follows:

\begin{enumerate}[(i)]
    \item Leakage predictions considering surface roughness \cite{persson2008theory}.
    \item Leakage predictions using different separation theories (critical junction and effective medium) \cite{Lorenz2009CJ, Lorenz2009EM}.
    \item Leakage of fluids in different flow regimes, diffusive or ballistic (gas) \cite{huon2022air}.
    \item Leakage with fluid-pressure-induced surface deformation \cite{lorenz2010time}.
    \item Influence of plastic deformation on the leak rate (in an approximate way) \cite{persson2016leakagemetal2, fischer2020metal1}. 
    \item Role of hydrophobicity on the leak rate \cite{lorenz2014role}.
\end{enumerate}

Although these studies have addressed many aspects, a common limitation is the assumption that the pressure distribution in the seal is either Hertzian (for O-rings) or rectangular (for square rings). These assumptions are not accurate in many practical applications.

%The present study specifically investigates the leakage of air at rubber-glass interfaces using a prototype of a syringe system (rubber stopper and glass barrel) from Beckton Dickinson (BD) Medical-Pharmaceutical Systems. However, the developed methodology establishes a generalized framework applicable to various contact systems. Nominal contact pressure distributions are determined using the Finite Element Method (FEM), and all relevant features are implemented into Multiscale Contact Mechanics (MCM) programs based on the Persson theory to predict leakage. The predictions are compared with experimental leakage tests, showing good agreement. Additionally, this study examines the dependency of leakage on contact pressure and material properties, particularly the elastic modulus $E$ and contact pressure $p_0$. This enhances the understanding of the correlation between leakage and contact conditions, providing valuable insights for improving the design of seals and mechanical systems.

%In the present study, the separation $u_{\rm c}$ at the critical junction is much smaller than the gas molecule mean free path $\lambda$, and the gas leakage occurs in the so-called ballistic limit (see \cite{huon2022air}). In Ref. \cite{huon2022air}, the leakage was analyzed using an approach in which the leak rate in the diffusive limit was scaled by a factor proportional to $\lambda/u_{\rm c}$ to estimate the leakage in the ballistic regime. In contrast, the present work employs leak rates calculated rigorously in the ballistic limit, as described in the Theory section and Appendix A.

The present study specifically investigates the leakage of air at rubber–glass interfaces using a prototype syringe system (rubber stopper and glass barrel) from Beckton Dickinson (BD) Medical-Pharmaceutical Systems. However, the developed methodology establishes a generalized framework applicable to a wide range of contact systems. Nominal contact pressure distributions, neglecting surface roughness, are determined using the Finite Element Method (FEM), and all relevant features are incorporated into Multiscale Contact Mechanics (MCM) software based on Persson theory to predict leakage.

In the system studied, the separation at the critical junction $u_{\rm c}$ is much smaller than the gas molecule mean free path $\lambda$, and the gas leakage occurs in the so-called ballistic limit (see \cite{huon2022air}). In Ref. \cite{huon2022air}, leakage was analyzed using an approach in which the leak rate in the diffusive limit was scaled by a factor proportional to $\lambda/u_{\rm c}$ to estimate leakage in the ballistic regime. In contrast, the present study employs leak rates calculated rigorously in the ballistic limit, as described in the Theory section and Appendix A.

The predicted leak rates are compared with experimental measurements, showing good agreement. Furthermore, this study examines the dependence of leakage on contact pressure and material properties, particularly the elastic modulus $E$ and nominal contact pressure $p_0$. These results enhance the understanding of the relationship between leakage and contact conditions, offering valuable insights for improving the design of seals and mechanical systems.

\begin{figure}[ht!]
    \centering
    \includegraphics[width=0.4\textwidth]{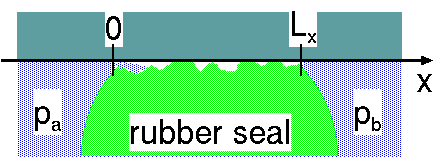}
    \caption{Cross-section of a circular seal orthogonal to the angular direction.
The fluid pressure $p_{\rm a} > p_{\rm b}$ and fluid leakage occurs along the positive $x$-axis.}
    \label{picSEAL.png}
\end{figure}

\section{Theory}

% where $p({\bf x},t)$ is the fluid pressure. 

Here we describe the theory used to obtain the leakage rate of fluids in torus-shaped rubber seals with an arbitrary cross section.
We calculate the fluid leakage using the effective medium theory developed in Ref. \cite{lorenz2010dependence} 
(see also Refs. \cite{Lorenz2009CJ, Lorenz2009EM, Persson2021comments, Wang2021, Scaraggi2015, Bottiglione2009}). 
Here we review the basic equations.

Consider the flow of a fluid at the interface between two solids with surface roughness.
For an incompressible fluid (liquid), the conserved quantity is the volume of fluid, and the continuity equation for fluid flow takes the form
$$\nabla \cdot {\bf J}_{\rm liquid} = {\partial u \over \partial t} \eqno(1)$$
where $u({\bf x},t)$ is the interfacial separation.
For a compressible fluid (gas), the conserved quantity is not the volume of fluid but the number of atoms (or molecules). In this case,
$$\nabla \cdot {\bf J}_{\rm gas} = {\partial n \over \partial t} \eqno(2)$$
where $n$ is the number of molecules per unit surface area. Note that the two flow currents have different dimensions: ${\rm m^2/s}$ for the incompressible fluid
and ${\rm 1/s m}$ for the (compressible) gas.

In what follows, we assume stationary conditions so that there is no dependency on time.
For an incompressible fluid, the (microscopic) flow current is related to the pressure gradient as follows \cite{Lorenz2009EM}:
$${\bf J}_{\rm liquid}({\bf x}) = -{u^3 \over 12 \eta} \nabla p. \eqno(3)$$
Here, $\eta$ is the fluid viscosity and $p({\bf x})$ is the pressure at the lateral point ${\bf x} = (x,y)$ at the interface. 
The pressure is assumed to be independent of the coordinate $z$, which is normal to the interface. 
The current ${\bf J}_{\rm liquid}$ represents the volume of fluid per unit surface area multiplied by the average flow velocity across the thickness of the gap 
(with units of ${\rm m^3/m^2 \times m/s = m^2/s}$). 
For a gas \cite{huon2022air}:
$${\bf J}_{\rm gas}({\bf x}) = -\left( {u^3 p \over 12 k_{\rm B} T \eta} + {{\bar v} u^2 \over 2 k_{\rm B} T} \right) \nabla p . \eqno(4)$$
Here we assume the temperature is constant. In (4) we interpolate between diffusive airflow (first term) and ballistic airflow (second term), see Ref. \cite{huon2022air}. That is, when the separation between the surfaces is much larger than the mean free path of gas molecules, the fluid flow is diffusive. In the opposite limit, there are no collisions between molecules, and they propagate ballistically between collisions with solid walls.

The flow conductivity $\sigma ({\bf x})$ is defined by 
$${\bf J_{fluid}} ({\bf x}) = - \sigma ({\bf x}) \nabla p ({\bf x})$$
so that
$$\sigma = {u^3 \over 12 \eta} \eqno(5)$$
for an incompressible liquid, and
$$\sigma = {u^3 p \over 12 k_{\rm B} T \eta} + {{\bar v} u^2 \over 2 k_{\rm B} T} \eqno(6)$$ 
for a compressible gas. For surfaces with roughness on many length scales, $u({\bf x})$, $p ({\bf x})$, and $\sigma ({\bf x})$ vary rapidly with the spatial coordinate ${\bf x}$.

Consider a torus-like seal (see Fig. \ref{picSEAL.png}) and let $x$ be the axis along the fluid leakage direction. 
The seal contacts a smooth surface for $0 < x < L_x$. Assume that the fluid pressure is $p_{\rm a}$ for $x < 0$ and $p_{\rm b} < p_{\rm a}$ for $x > L_x$. In this case, fluid flow occurs in the positive $x$-direction.

For an incompressible fluid, $\sigma$ is given by (5) and depends on the spatial coordinate ${\bf x}$ only via $u$. For a compressible gas, $\sigma$ is given by (6) depends on $u({\bf x})$
{\it and} the fluid pressure $p({\bf x})$. This makes the problem more complex. To simplify, we replace $p$ in (6) with $(p_{\rm a} + p_{\rm b})/2$, i.e., the average of the fluid pressure between the two sides of the seal region. After this approximation, $\sigma$ depends on ${\bf x}$ only via $u({\bf x})$.

The ensemble-averaged fluid flow current is
$${\bf J} = - \sigma_{\rm eff} \nabla p_{\rm fluid}$$
where $p_{\rm fluid}$ is the ensemble-averaged fluid pressure and $\sigma_{\rm eff}$ is the corresponding 
effective flow conductivity. Here ${\bf J}$, $\sigma_{\rm eff}$, and $p_{\rm fluid}$ vary slowly in space and 
are obtained from the corresponding microscopic quantities in equations (1)-(6) 
after averaging out the roughness (ensemble-averaging). 
Thus, while ${\bf J}_{\rm fluid} ({\bf x})$ varies rapidly in space due to surface roughness, 
the ensemble-averaged current ${\bf J} = \langle {\bf J}_{\rm fluid} \rangle$ represents an effective flow current for a smooth surface. 
This averaged current varies spatially only due to the macroscopic shape of the contact. 
The averaging is carried out using the Bruggeman effective medium theory (see \cite{dapp2012self,Bruggeman1935}, section below, and Appendix A, where a effective flow conductivity $\sigma_{\rm eff}$ is calculated).

The probability distribution of interfacial separation can be written as
$$P(u) = {A \over A_0} \delta(u) + P_{\rm c}(u) \eqno(7)$$
where $A/A_0$ is the relative contact area ($u=0$ contribution to $P(u)$) and where $\delta(u)$ is the Dirac delta function which vanish for $u\neq0$. 
$P_{\rm c}(u)$ is the continuous part of the distribution and is non-zero for $u > 0$. 
In the effective medium approximation, the fluid flow conductivity is given by
$${1 \over \sigma_{\rm eff}} = {2 \over \sigma_{\rm eff}} {A \over A_0} + \int_0^\infty du \, {2P_{\rm c}(u) \over \sigma + \sigma_{\rm eff}} \eqno(8)$$
where $\sigma$ is given by (5) for an incompressible fluid, and by (6) for a compressible gas. 
Note that $\sigma_{\rm eff}$ occurs on both sides of (8) and (8) must be solved by iteration.
Eq. (8) assumes isotropic roughness, but the theory can be extended to anisotropic roughness \cite{Persson2020interfacial, Persson2021comments, Wang2021} (see also Appendix A). 

When the contact area percolate no fluid can flow at the interface which correspond to $\sigma_{\rm eff}=0$. Multiplying
(8) with $\sigma_{\rm eff}$ and taking the limit  $\sigma_{\rm eff} \rightarrow 0$ gives $1= 2A/A_0$ or $A/A_0 = 0.5$.
Thus (8) predicts that the contact area percolates at $A/A_0 = 0.5$, but simulations show that for random roughness, percolation occurs at $A/A_0 \approx 0.42$
(see Ref. \cite{dapp2012self}). In Refs. \cite{dapp2012self, Persson2020interfacial, Persson2021comments, Wang2021} (see also Appendix A), it was shown how to modify the Bruggeman theory to give the correct percolation threshold \footnote{The equation for how to obtain the correct percolation limit given in Ref. \cite{Persson2020interfacial} contained a misprint which was pointed out by Wang et al (Ref. \cite{Wang2021}) and was corrected in Ref. \cite{Persson2021comments}.}.

In (8), the probability distribution $P(u)$ is calculated using the Persson contact mechanics theory. 
For the torus-like geometry relevant here, 
the ensemble averaged fluid pressure $p_{\rm fluid}(x)$ and contact pressure $p_{\rm cont}(x)$ (the part of the total
pressure resulting from the asperity contact area $A$) depend only on $x$. 

In this case, for stationary flow, the continuity equation $\nabla \cdot {\bf J} = 0$ becomes
$${d J_x \over dx } = {d \over dx} \left( \sigma_{\rm eff} {d p \over dx} \right) = 0$$
This equation is easy to integrate \cite{lorenz2010time}:
Let
$$S(x) = \int_0^x dx' \ \sigma_{\rm eff}^{-1} (p_{\rm cont}(x')) \eqno(9)$$
then the leak rate is
$$\dot Q = {L_y \over S(L_x)} (p_{\rm a} - p_{\rm b}) \eqno(10)$$
where $L_y = 2\pi R$ is the length of the O-ring in the angular direction.
For an incompressible fluid the leak rate $\dot Q$ is the volume of fluid per unit time ($\rm m^3/s$)
and for a compressible gas, it is the number of molecules per unit time ($\rm 1/s$).

The fluid pressure is
$$p_{\rm fluid}(x) = p_{\rm a} - (p_{\rm a} - p_{\rm b}) {S(x) \over S(L_x)} \eqno(11)$$

The ensemble-averaged pressure acting on the rubber surface is
$$p_0(x) = p_{\rm cont}(x) + p_{\rm fluid}(x) \eqno(12)$$
The pressure $p_0(x)$ arises from macroscopic deformation of the rubber and was obtained from FEM for the rubber stopper assuming smooth contact surfaces. The small change in interfacial separation induced by the fluid pressure has a negligible effect on $p_0(x)$, except near $x = 0$, where $p_0(x) \rightarrow 0$. In this region, lift-off occurs. Where the local surface separation exceeds the characteristic surface roughness, the rubber surface is subjected to the fluid pressure $p_{\rm a}$; elsewhere, the pressure is given by equation (11).

\section{Leakage Prediction Using Persson Contact Mechanics Software}

\begin{figure}[ht!]
    \centering
    \includegraphics[width=0.4\textwidth]{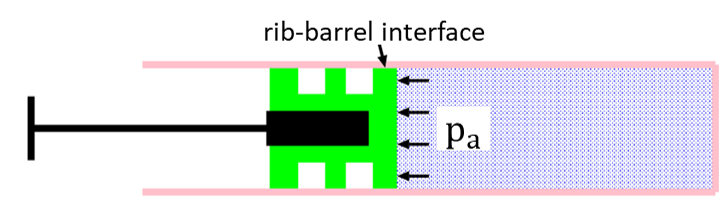}
    \caption{Schematic representation of the studied system.}
    \label{system}
\end{figure}

A schematic representation of the system used in this study is shown in Fig. \ref{system}. The rubber stopper features three ribs with different shapes and heights that come into contact with the inner wall of the barrel. The present study focuses on the first rib, which is located closest to the fluid inside the barrel, since it plays the most critical role in preventing leakage. The aim is to assess the sealing performance of this specific interface. Therefore, this study concentrates on the interface between the first rib and the barrel, as indicated by the black arrow in the figure.

\begin{figure*}[ht!]
    \centering
    \includegraphics[width=0.4\textwidth]{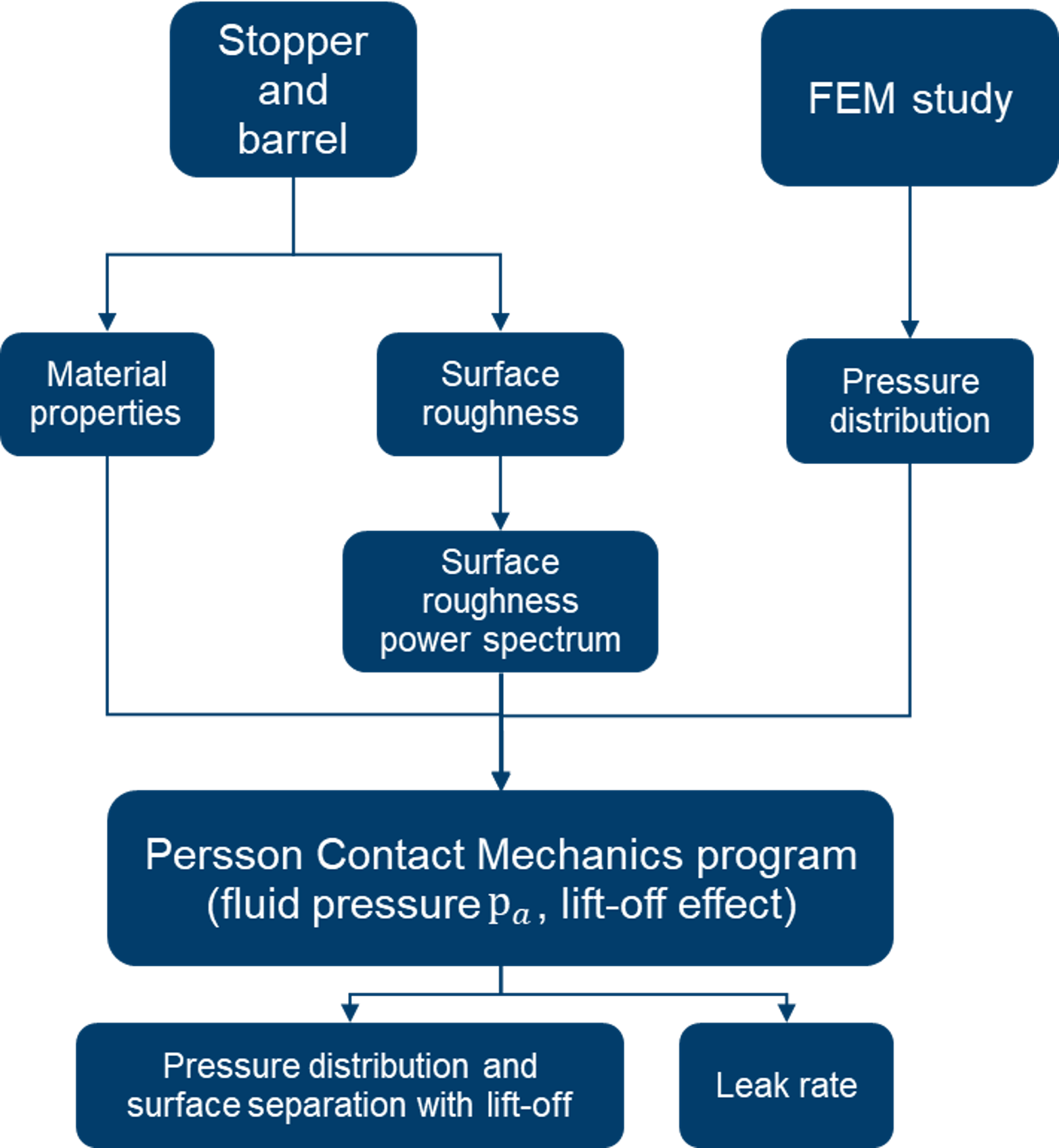}
    \caption{General process for investigating leakage using MCM software.}
    \label{process}
\end{figure*}

The general process for leakage prediction using MCM software is illustrated in Fig. \ref{process}.

Among the inputs required for the theoretical calculations, the surface roughness power spectral density (PSD) is the most critical. It is typically obtained from measured topography data using stylus or atomic force microscopy (AFM) techniques. Another key input is the pressure distribution within the rib-barrel interface. In this study, due to the complex geometry, the pressure distribution is non-symmetric and was determined using FEM analysis. Additional input parameters include the mechanical and rheological properties of the solid and fluid materials, such as the elastic modulus, Poisson’s ratio, and viscosity.

With all inputs prepared, the MCM software calculate important interfacial quantities such as contact pressure, surface separation, and leak rate, while accounting for the effect of lift-off (fluid-induced elastic deformation of the stopper).

\subsection{Surface Topography Power Spectra}

We measured the topography of the rubber surface (on the first rib) using two engineering stylus instruments and an AFM. These methods have been shown to effectively characterize surface roughness at different length scales \cite{Rodriguez2025}. The surface of the glass tube is assumed to be smooth, as its roughness is significantly lower than that of the rubber surface.

Two sets of engineering stylus measurements were performed: (a) using a Mitutoyo Portable Surface Roughness Measurement Surftest SJ-410 equipped with a diamond tip having a radius of curvature $R = 1 \ {\rm \mu m}$ and a tip-substrate repulsive force $F_N = 0.75 \ {\rm mN}$. The step length (pixel size) was $0.5 \ {\rm \mu m}$, the scan length was $L = 2 \ {\rm mm}$, and the tip speed was $v = 50 \ {\rm \mu m/s}$; (b) using a Bruker Dektak XT, equipped with a diamond tip of radius $R = 0.7 \ {\rm \mu m}$ and a tip-substrate repulsive force $F_N = 1 \times 10^{-5} \ {\rm N}$. The scan length was $L = 3 \ {\rm mm}$ with a step resolution of $0.139 \ {\rm \mu m}$. The tip speed ranged from $33$ to $44 \ {\rm \mu m/s}$. The Bruker Dektak was mounted on a vibration isolation table.

AFM measurements were conducted using a Bruker Dimension 3100 in tapping mode (amplitude modulation), equipped with an RSTESPA-300 probe having a tip radius of $8 \ {\rm nm}$. The two-dimensional scan resolution was $1024 \times 1024$ pixels over an area of $20 \ {\rm \mu m} \times 20 \ {\rm \mu m}$.

\begin{figure}[ht!]
    \centering
    \includegraphics[width=1\columnwidth]{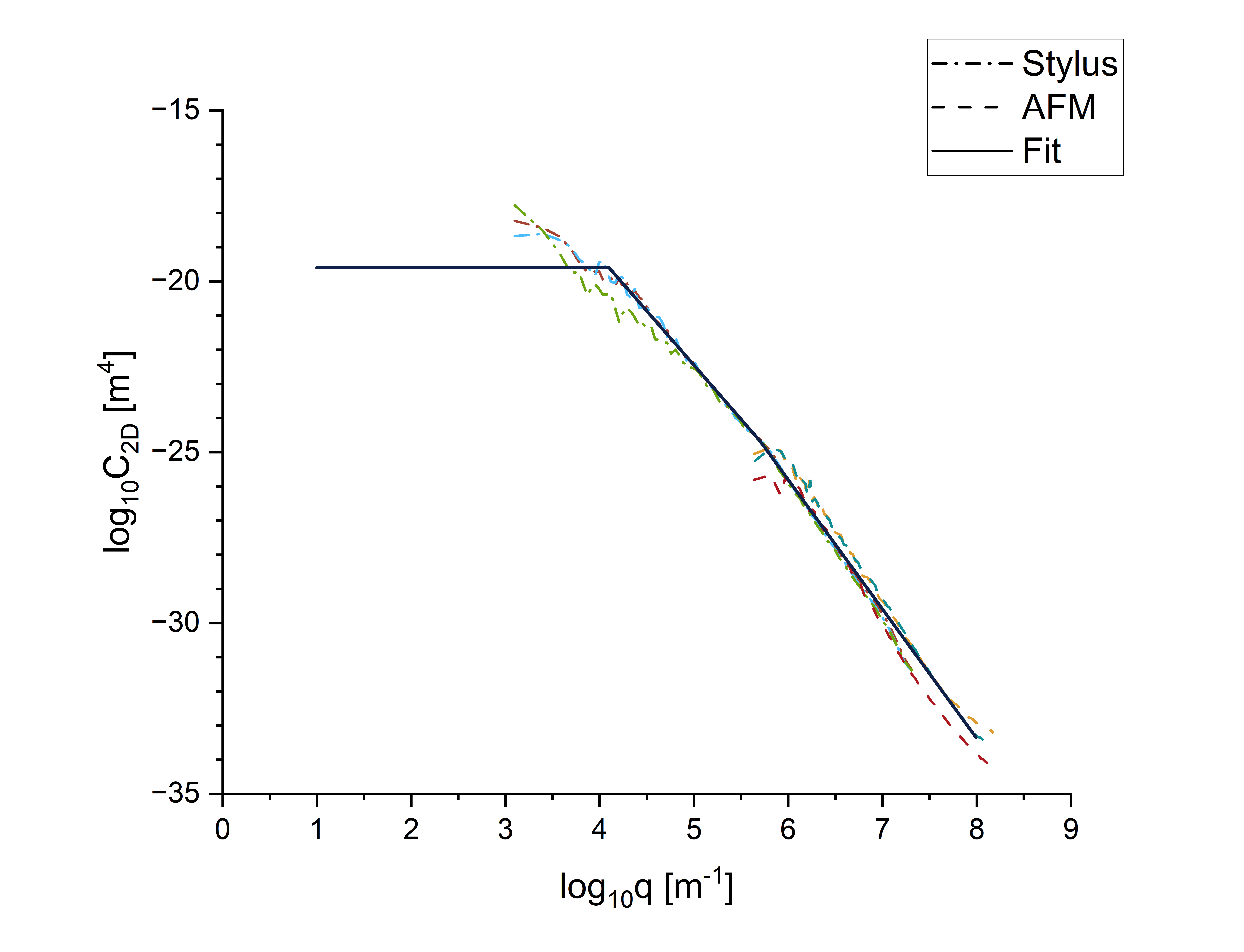}
    \caption{2D surface roughness power spectra calculated from stylus (dash-dot lines), AFM (dashed lines), and fitted PSD (solid line) used in calculations.}
    \label{PSD}
\end{figure}

The topography data were processed using the MCM software to calculate the two-dimensional surface roughness PSD. The results are shown in Fig. \ref{PSD}. It can be seen that the stylus measurements (dash-dotted lines) and the AFM measurements (dashed lines) overlap well. Based on these data, a fitted PSD curve (solid line) was used for subsequent calculations. Note that, for technical reasons, a roll-off region was added to the fitted PSD at the wavenumber $q_{\rm r} = 2\pi/L$, where $L$ is the width of the contact region in the flow direction (i.e., the $x$-direction).

\subsection{Contact Pressures}

\begin{figure}[ht!]
    \centering
    \includegraphics[width=1\columnwidth]{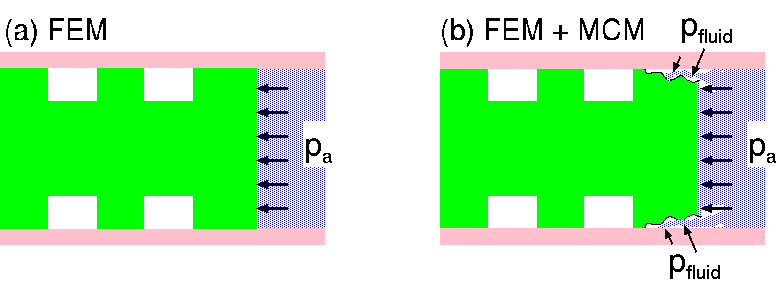}
    \caption{(a) During fluid injection, the fluid pressure in the syringe compresses the rubber stopper, increasing the contact pressure at the rubber-barrel interface. This effect is included in the FEM analysis. (b) If open regions at the rubber-barrel interface are connected to the fluid via open channels, the fluid pressure will increase the average surface separation at the interface.}
    \label{SchematicContact.png}
\end{figure}

The contact pressure in the rib-barrel interface was determined using FEM simulations. The influence of fluid pressure $p_{\rm a}$ on the contact pressure was modeled by applying pressure to the mesh at the stopper-fluid interface [see Fig. \ref{SchematicContact.png}(a)]. Seven distinct values of $p_{\rm a}$ were considered: $0,\ 0.069,\ 0.138,\ 0.207,\ 0.276,\ 0.345$, and $0.414 \ {\rm MPa}$ (corresponding to $0$–$60$ psi in imperial units).

In the FEM simulation, the stopper was slightly tilted relative to the barrel, resulting in non-uniform contact pressures along different axial paths parallel to the barrel axis. Two axial paths representing the nominal highest and lowest pressures were selected for analysis. These paths serve as upper and lower bounds for the leakage rate.

\begin{figure}[ht!]
    \centering
    \includegraphics[width=1\columnwidth]{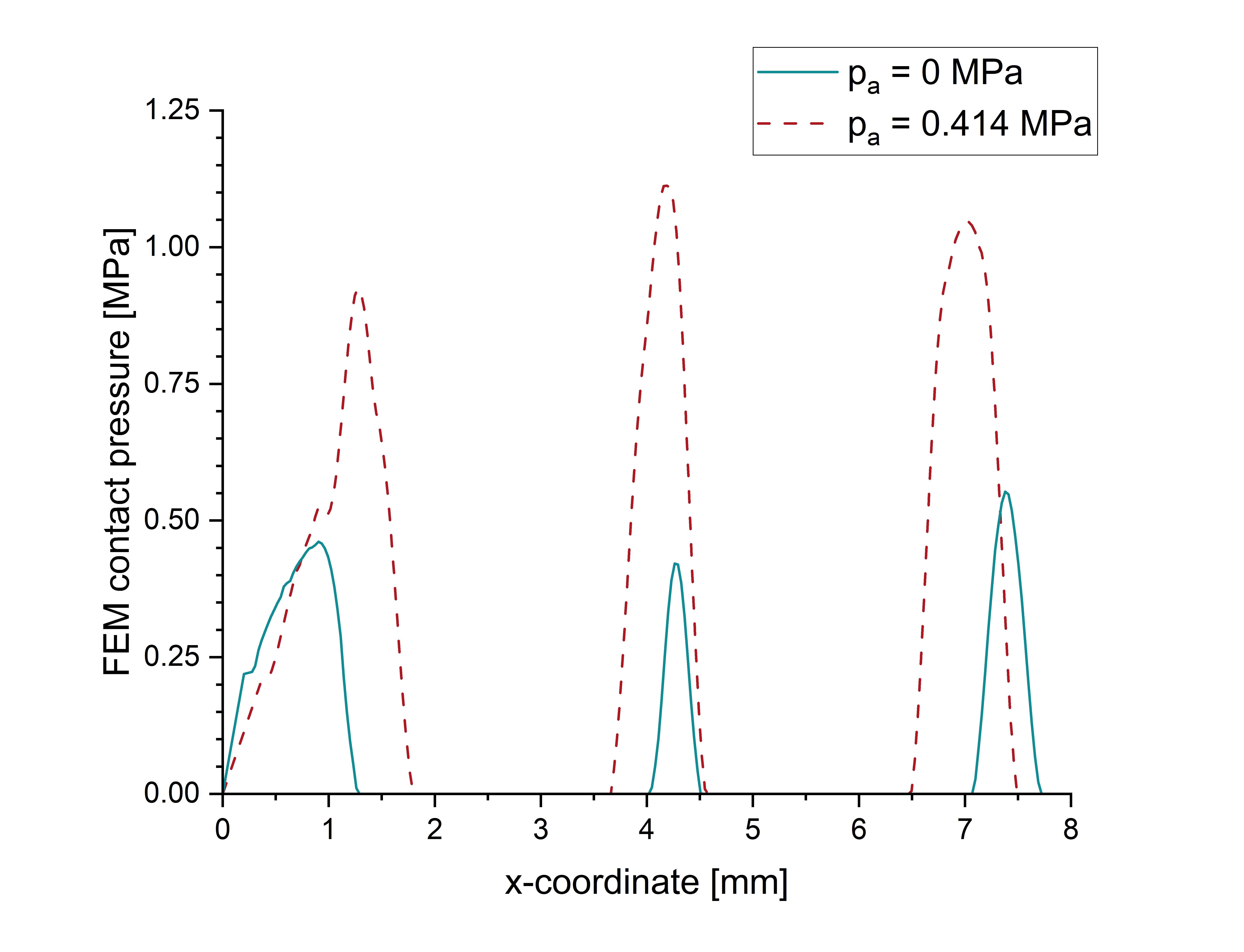}
    \caption{FEM-simulated contact pressure distributions $p_0(x)$ along the lowest pressure path at fluid pressures $p_{\rm a}$ of $0$ and $0.414\ {\rm MPa}$. The three peaks represent the three ribs of the stopper. Only the pressure distribution of the first rib (the leftmost peak) is used in the simulation.}
    \label{contact_pressure.png}
\end{figure}

Fig. \ref{contact_pressure.png} shows FEM-derived contact pressure distributions along the lowest pressure path at $p_{\rm a} = 0$ and $0.414 \ {\rm MPa}$. The three distributions from left to right correspond to the three ribs on the stopper, with the stopper-fluid interface located on the far left. In the simulation, only the contact pressure distribution corresponding to the first rib (the first peak on the left) is used. It can be observed that applying fluid pressure to the stopper increases the contact pressure through elastic deformation.

In practice, the rubber stopper does not make full contact with the barrel due to surface roughness and the relatively small applied load. As a result, non-contact regions are present at the interface. If these regions are connected to the pressurized fluid inside the stopper, the fluid pressure at the interface will increase the average surface separation [see Fig. \ref{SchematicContact.png}(b)]. This phenomenon is known as ``lift-off". The MCM software calculations presented below incorporate this effect accurately.

\subsection{Other Material Properties}

The stress-strain behavior of rubber materials is generally nonlinear, and this is accounted for in the FEM analysis. However, Persson contact mechanics assumes linear stress-strain behavior. One approach to obtain an effective modulus $E_{\rm eff}$ for use in Persson theory is to estimate the strain in the asperity contact regions and use the secant modulus corresponding to that strain from measured stress-strain curves in tension or compression. 

Another approach, which we adopt here, is to fit the contact pressure distribution obtained from FEM to the distribution predicted by Hertzian contact theory, applied to the place where the contact is approximately Hertz-like. Assuming the material was linearly elastic, the Hertz theory would apply and an effective elastic modulus $E_{\rm eff}$ can be driven from the fit. This method has the advantage of producing a strain field that closely resembles that of asperity contact, involving not only compressive strains but also some shear deformation (included in the pressure distribution from FEM).

Thus, we determine the effective modulus $E_{\rm eff}$ by ensuring that the calculated Hertz contact pressure distribution closely matches the FEM results obtained using a full nonlinear material model. In this study, the Yeoh rheological model was used in the FEM simulations \cite{Yeoh_1993}. The second rib is approximated as having a half-cylinder cross-section, and the rib-barrel contact is assumed to follow Hertzian contact mechanics. From the radius of curvature of the rib and the contact width obtained from FEM, we chose $E_{\rm eff}$ to reproduce the maximum contact pressure predicted by FEM. This yields an effective modulus of $E_{\rm eff} \approx 1.8 \ {\rm MPa}$. We estimate that the compression of this rib ($\sim 35 \%$) is also similar to the compression of the asperity contact regions, therefore, the result can be used on the first rib.

This study investigates the leakage of air. To calculate the leak rate, both the viscosity and the effective mean free path of air molecules are required. If the mean free path is much larger than the separation between surfaces at the narrowest constrictions (known as critical junctions) along the gas flow channels, the gas flow is ballistic, and molecules will only collide with the wall. In the opposite limit, the motion is diffusive. The formalism used in this study accounts for both cases by interpolating between the diffusive and ballistic regimes \cite{huon2022air}. 

Here, the viscosity of air is taken as $\eta_{\rm air} \approx 10^{-6} \ {\rm Pa \cdot s}$, and the mean free path is chosen as $60 \ {\rm nm}$.

\section{Experimental Setup}

\begin{figure*}[ht!]
    \centering
    \includegraphics[width=0.6\textwidth]{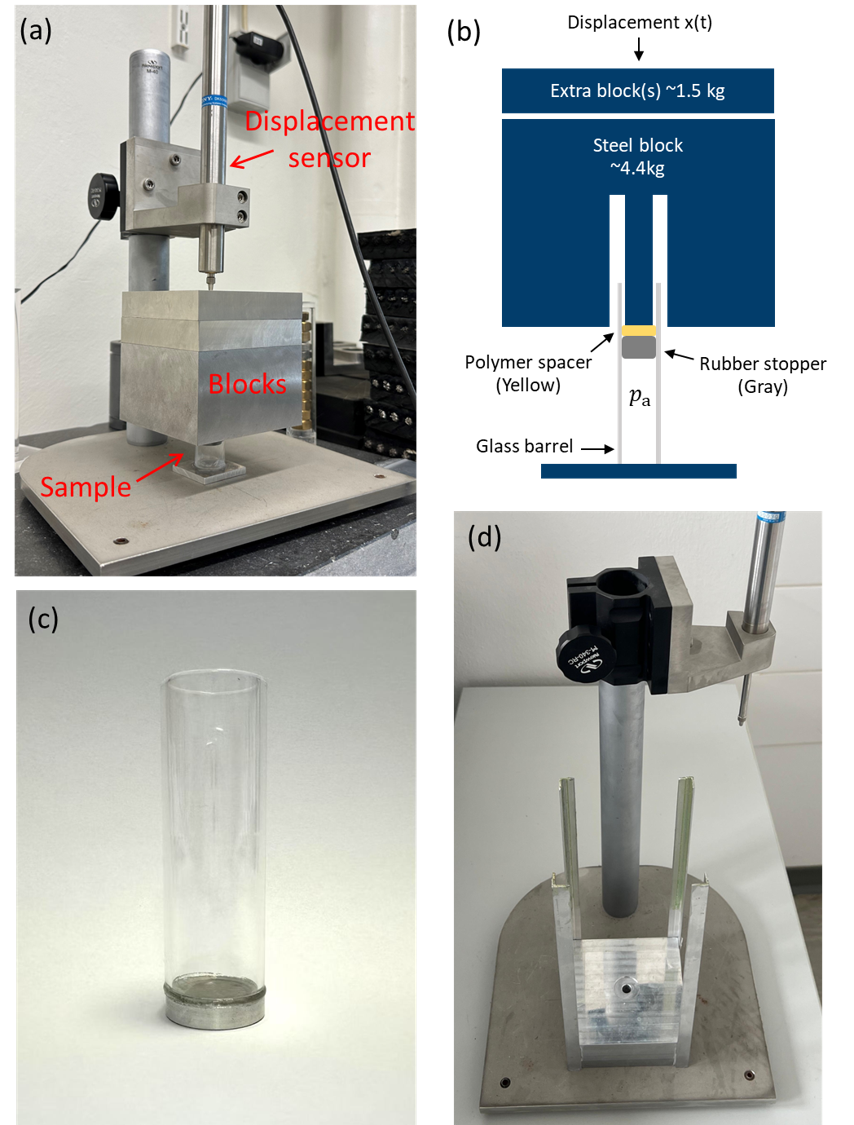}
    \caption{(a) Experimental setup used in this study. (b) Schematic representation of the setup. (c) Glass barrel glued to a circular aluminum plate. (d) Base with rails on all four corners and a central recess to ensure proper alignment of the barrel and blocks.}
    \label{exp_setup}
\end{figure*}

The test rig is shown in Fig. \ref{exp_setup}(a). It consists of several steel blocks and a Sony DK50NR5 displacement sensor with a resolution of $0.5 \ {\rm \mu m}$. The largest block weighs approximately $4.4 \ {\rm kg}$, and additional weights can be added to achieve different values of $p_{\rm a}$. This method enables the consistent application of large loads, with a practical upper limit of approximately $20 \ {\rm kg}$ when using lead blocks as the extra loading mass. The largest block includes a milled cylinder that pushes the rubber stopper into the glass barrel [see Fig. \ref{exp_setup}(b)]. During testing, a polymer spacer resembling a syringe piston is used to maintain proper compression and tension in the stopper, ensuring correct movement and avoiding tilting. The glass barrel of the prototype used for testing comes from the same production line as actual syringe barrels, with the only difference being its shape. One end of the glass barrel is glued to a circular aluminum plate to ensure a tight seal [see Fig. \ref{exp_setup}(c)]. Finally, the barrel is positioned at the center of a base equipped with guiding rails at the corners [see Fig. \ref{exp_setup}(d)]. These rails guide the motion of the steel block, ensuring that the cylinder in the center of the block aligns precisely with the center of the glass barrel and pushes the stopper without contacting the barrel walls.

The basic principle of the setup is that by sealing the lower end of the barrel and applying a load to the stopper, any leakage is reflected by the displacement of the steel block over time under load. The volume of leaked air can be calculated by multiplying the measured displacement by the cross-sectional area of the barrel. This simple test method has been shown to be effective and provides results consistent with more complex techniques, such as helium leak testing \cite{huon2022air}.

\begin{figure}[ht!]
    \centering
    \includegraphics[width=1\columnwidth]{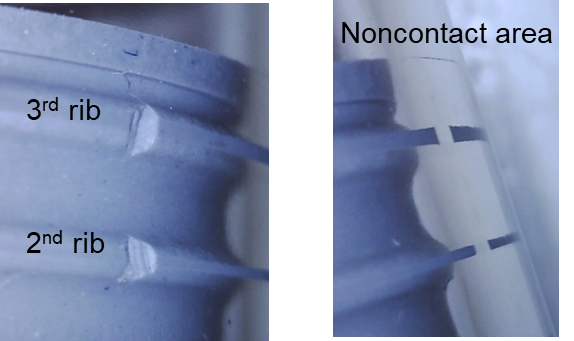}
    \caption{Cuts and corresponding non-contact areas on the second and third ribs. The cuts were not large enough to significantly alter the contact pressure but were sufficient to allow air to pass through.}
    \label{cut_and_path}
\end{figure}

Experimental measurements of air leakage were conducted using standard-size stoppers and glass barrels with an inner diameter of $d = 19.05 \ {\rm mm}$. To match the simulation conditions, the experiments focused solely on leakage occurring at the first rib. To ensure this, cuts were made using a scalpel on the second and third ribs of the stopper, as shown in Fig. \ref{cut_and_path}. These cuts were large enough to allow fluid (air) to pass through but not large enough to significantly affect the contact pressure.

The stoppers and barrels were siliconized at the factory. Therefore, leakage tests involving silicone oil were conducted first. To better match the simulation conditions (i.e., dry contact between the stopper and barrel), subsequent tests were carried out after thoroughly cleaning both components with isopropyl alcohol to remove most of the silicone oil.

A key challenge in this experimental setup is estimating the internal pressure in the barrel, $p_{\rm a}$, which is calculated as $p_{\rm a} = F_N / A_0$, where $F_N$ is the applied normal force and $A_0$ is the inner cross-sectional area of the barrel. Friction between the stopper and barrel, as well as between the steel block and guiding rails, may alter the actual value of $F_N$. In practice, some approximations are made: the guiding rails are lubricated with grease to minimize friction, which is then considered negligible. The friction between the stopper and barrel is measured to be on the order of a few newtons in the siliconized case and approximately $10 \ {\rm N}$ under dry conditions. In both cases, the friction force is small compared to the applied load ($\geq 44 \ {\rm N}$).

In total, approximately 20 randomly selected stoppers and 10 randomly selected glass barrels were tested. The combinations of stoppers and barrels were assembled in a standard (non-cleanroom) environment. After each trial, a visual inspection was performed to assess the condition of the components. This is important because contamination during assembly (e.g., dust particles or glass fragments) may alter the contact conditions and lead to anomalous test results, typically characterized by high leakage rates. In rare cases, such anomalies may also be due to defects in the stopper itself. If the glass barrel was determined to be in good condition, the stopper was replaced and the experiment was repeated. If not, a new barrel was used. The old barrel and, in some cases, the stoppers were cleaned with isopropyl alcohol and reused for non-siliconized tests.

In summary, all barrels and most stoppers underwent experiments with and without silicone oil. However, this does not imply that every stopper was tested in every barrel, as the combinations were relatively random.

\section{Comparison Between Predictions and Experimental Results}

Seven combinations of blocks were used in the test, resulting in $p_{\rm a}$ values ranging from $0.15$ to $0.5 \ {\rm MPa}$. Note that $0.15 \ {\rm MPa}$ corresponds to using only the largest block, which also defines the minimum $p_{\rm a}$ achievable in our setup.

\begin{figure}[ht!]
    \centering
    \includegraphics[width=1\columnwidth]{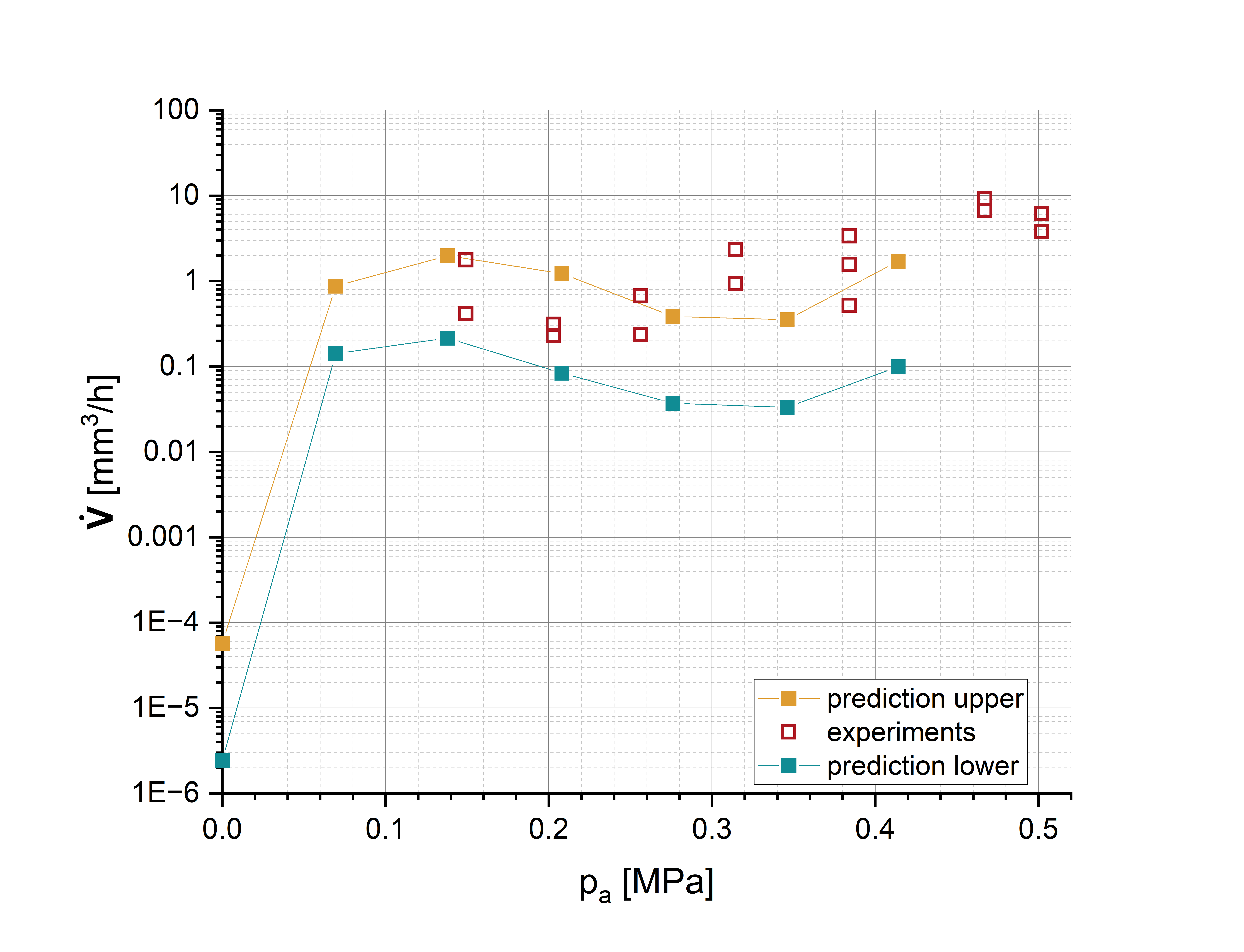}
    \caption{Leak rate as a function of $p_{\rm a}$. Comparison between simulation predictions (solid squares and lines) and experimental measurements (hollow squares) for clean syringes (nominally without silicone oil).}
    \label{exp_leak_rate}
\end{figure}

For syringes with silicone oil, no leakage was observed. For nominally oil-free systems, leakage was detected and fell within the predicted range, as shown in Fig. \ref{exp_leak_rate}. Both predicted and measured leak rates show an increasing trend (although non-monotonic) with fluid pressure $p_{\rm a}$, which is consistent with the analysis of how lift-off influences the leak path size and resulting leak rate (discussed below).

The absence of leakage in siliconized systems can be attributed to the formation of capillary bridges and/or the time required to squeeze out the high-viscosity silicone oil. If silicone oil forms capillary bridges across the narrowest constrictions along the leakage flow paths, leakage can only occur if the fluid pressure $p_{\rm a}$ is sufficiently high to break these bridges. Given the good wetting properties of silicone oil for both rubber and glass surfaces, capillary bridges are expected to form when the surface separation is small.

The pressure required to break a capillary bridge is given by the Laplace pressure $\Delta p$, determined by the Young–Laplace equation \cite{adamson1967physical, kralchevsky2001particles}:
$$\Delta p = \gamma \left( \frac{1}{R_1} + \frac{1}{R_2} \right)$$
where $R_1$ and $R_2$ are the principal radii of curvature of the capillary bridge. In our experiments, the leaking fluid is air, and $\gamma$ is the surface tension of silicone oil. If the syringe is filled with water, then $\gamma$ refers to the surface tension of the silicone oil-water interface. Typically, $R_2 \gg R_1$, and $R_1 \approx u_c/2$, where $u_c$ is the height of the critical junction. Therefore,
$$\Delta p \approx \frac{2\gamma}{u_c}$$
This relation implies that the pressure required to remove the capillary bridge is inversely proportional to the surface separation at the critical junctions. Hence, squeeze-out of silicone oil occurs only when $p_{\rm a} > \Delta p$. For lower pressures, no leakage occurs if sufficient silicone oil exists to form capillary bridges.

More generally, if no leakage is observed within a given time period, it suggests either that $p_{\rm a}$ is too low to break the capillary bridges, or that the volume of silicone oil is sufficiently large that it is not squeezed out within the observation period. The time required for a complete squeeze-out depends on the silicone oil's viscosity, the pressure difference, and the length of the leak path. In siliconized systems, the presence of substantial amounts of silicone oil can significantly delay extrusion, preventing observable leakage during the test duration. Note that the viscosity of silicone oil ($1 \ {\rm Pa \cdot s}$) is approximately $10^6$ times higher than that of air.

The strength of capillary bridges in a siliconized system is influenced by a complex interplay of factors. As $p_{\rm a}$ increases, the contact pressure also rises due to stopper deformation. While the increased contact pressure tends to reduce the surface separation and thus strengthen the capillary bridge, the fluid pressure lift-off effect acts in the opposite direction.

To remove silicone oil, some stoppers and barrels were cleaned with isopropyl alcohol. However, even after cleaning, traces of silicone oil may remain on the surface, or silicone oil present within the rubber may slowly diffuse back to the surface. Although the residual film thickness is likely on the nanometer scale, the separation at the critical junctions is also on the order of a few to tens of nanometers. If sufficiently small, these separations may still allow the formation of capillary bridges by residual silicone oil.

Another factor influencing leakage is the viscoelastic relaxation of the stopper material. Upon loading into the barrel, the stopper relaxes over time and becomes elastically softer (i.e., the effective modulus $E$ decreases). This reduces the contact pressure but, at the same time, the softer rubber can more easily conform to the surface roughness. For a linearly elastic material, these two effects exactly compensate each other (this will be discussed later), and the leak rate depends on the ratio $p_0/E$. 

If the dominant surface roughness is on the barrel side (which is usually not the case), then during motion, the asperity contact regions on the rubber side are continuously renewed. In this case, the rubber near the asperities remains elastically stiff, while it is soft on the macroscopic rib scale. As a result, the leak rate may increase with the time of stationary contact before sliding begins.

%%%%%%%%%%%%%%%%

\section{Elastic Modulus and Pressure Sensitivity Test}

The sealing process is sensitive to the contact conditions, particularly the contact pressure $p_0$ and the elastic modulus $E$. When $E$ is fixed, theory predicts that 
higher $p_0$ leads to smaller surface separation $u$, and thus a lower leak rate. Varying $E$ gives a similar effect: under the same contact pressure, a higher $E$ results in smaller asperity deformation, leading to a larger leak path and higher leak rate. This aligns with the common understanding that softer materials provide better sealing. 

The dependency on $p_0$ and $E$ becomes more critical when the contact approaches the percolation threshold, which depends on the stopper and barrel dimensions. 
To verify this, we performed a sensitivity test by varying $p_0$ and $E$ and examining 
their influence on leakage for the path with the lowest contact pressure and highest leak rate.

\begin{figure}[H]
    \centering
    \includegraphics[width=1\columnwidth]{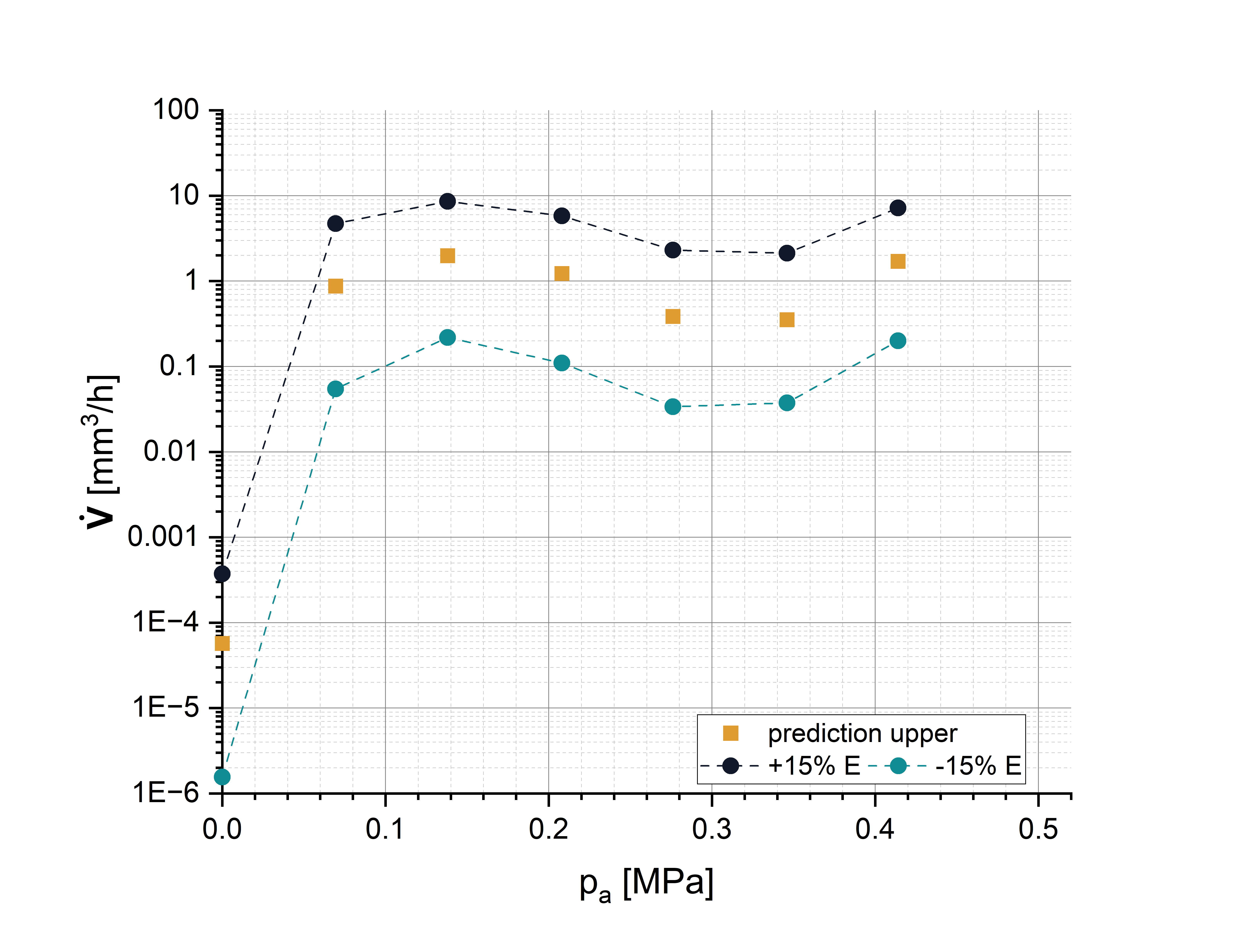}
    \caption{Predicted leak rate with $E$ varied by $\pm 15\%$ (blue and green dots with dashed lines).}
    \label{leak_rate_varE}
\end{figure}

We first used the original pressure distribution and varied the effective elastic modulus $E_{\rm eff} = 1.8 \ {\rm MPa}$ by $\pm 15\%$, resulting in $E^{+} = 2.07\ {\rm MPa}$ and $E^{-} = 1.53\ {\rm MPa}$. The results are shown in Fig. \ref{leak_rate_varE}.

\begin{figure}[H]
    \centering
    \includegraphics[width=1\columnwidth]{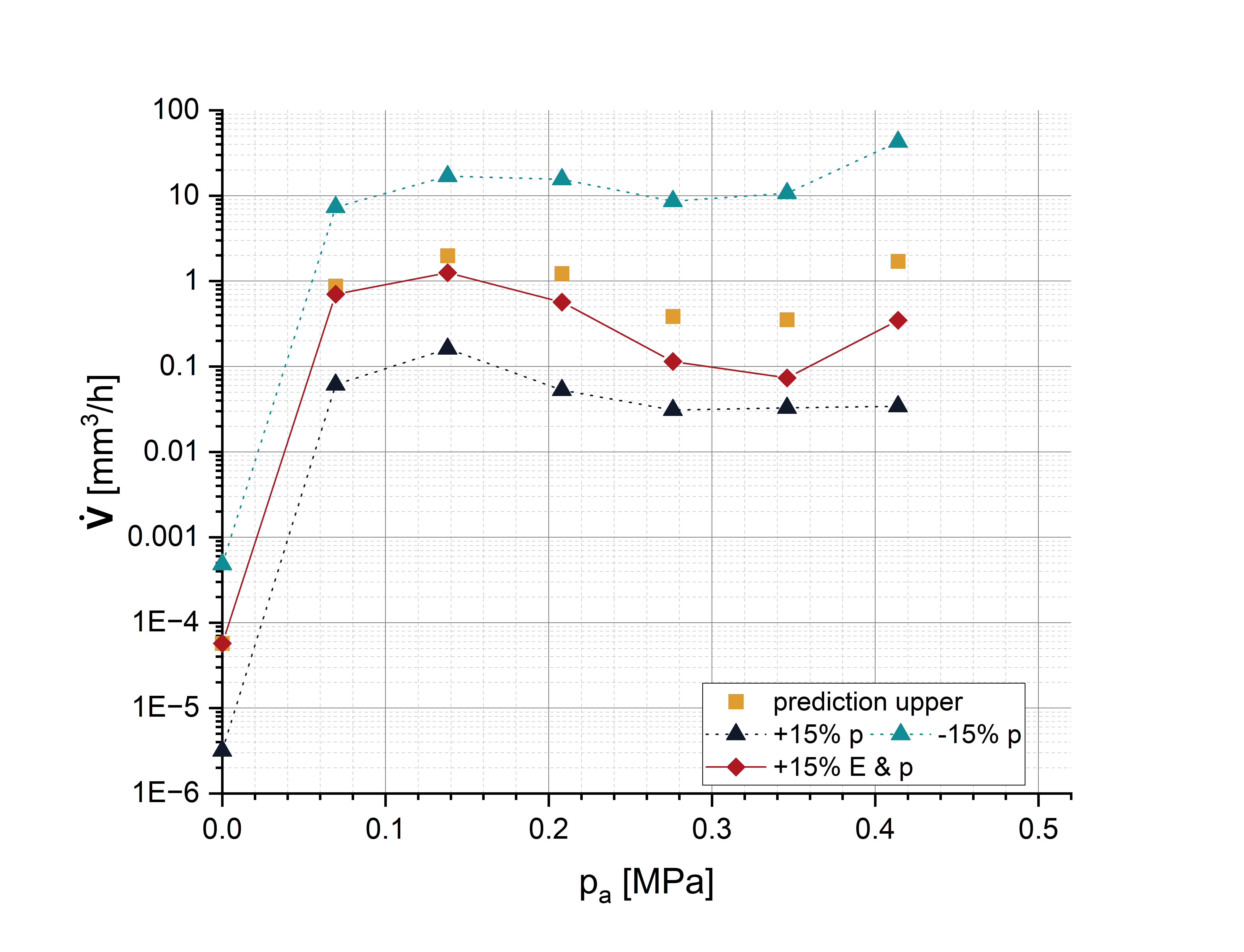}
    \caption{Predicted leak rate with $p_0$ varied by $\pm 15\%$ (blue and green triangles with dotted lines), and both $p_0$ and $E$ varied by $+15\%$ (red diamonds with line).}
    \label{leak_rate_varp}
\end{figure}

Next, using the original effective modulus $E_{\rm eff} = 1.8 \ {\rm MPa}$, we scaled the original contact pressure distribution 
$p_0(x)$ by $\pm 15\%$. These results are presented in Fig. \ref{leak_rate_varp}, along with an additional case where both $p_0$ and $E$ are increased by $15\%$.

\begin{figure}[H]
    \centering
    \includegraphics[width=1\columnwidth]{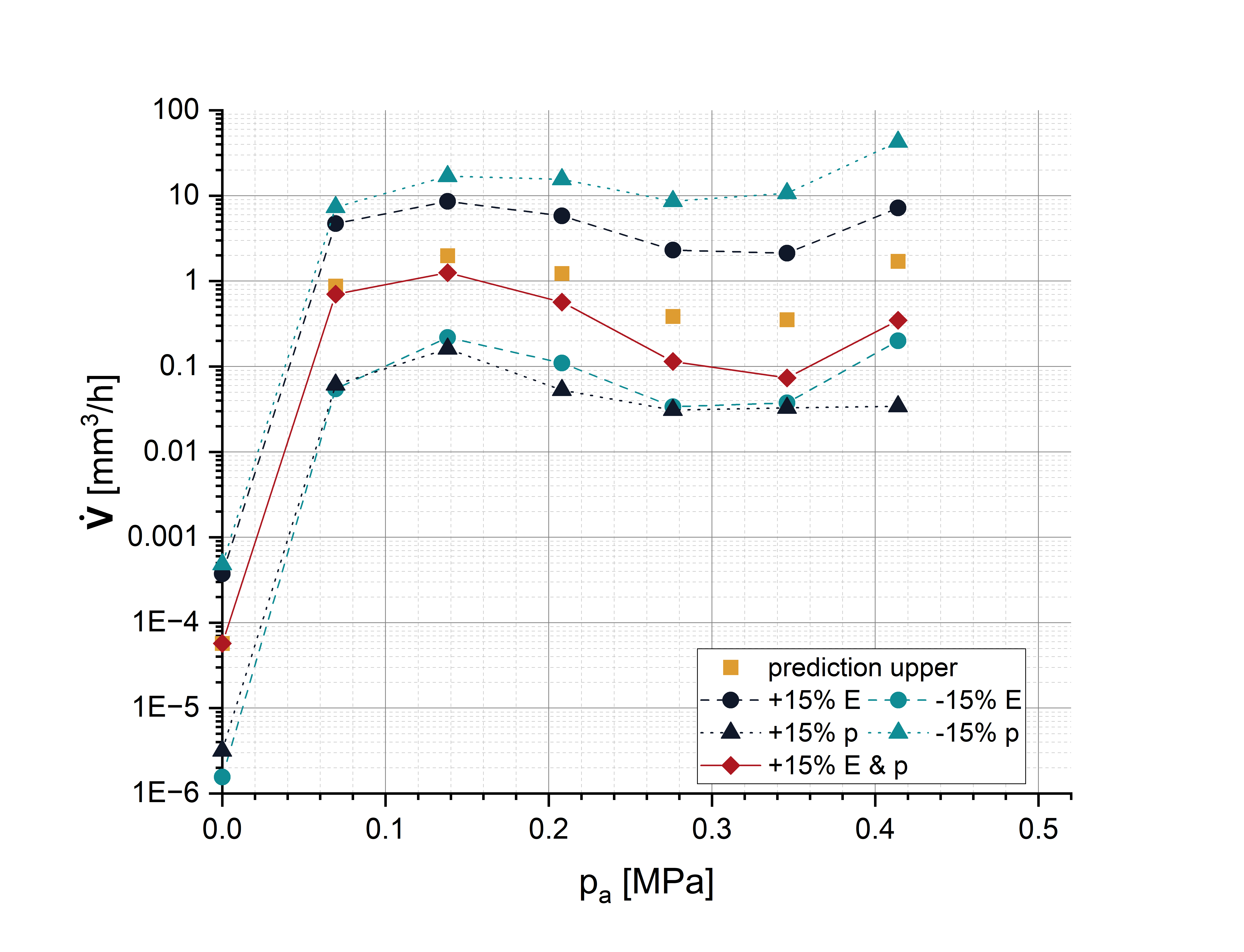}
    \caption{Predicted leak rate with variations in $p_0$ and/or $E$.}
    \label{leak_rate_varall}
\end{figure}

Fig. \ref{leak_rate_varall} summarizes all results from the sensitivity test.

From these results, we draw the following conclusions:

\begin{enumerate}
    \item When $p_0$ is fixed, the leak rate increases with increasing $E$.

    \item When $E$ is fixed, the leak rate decreases with increasing $p_0$.

    \item There is a nonlinear relationship between the leak rate $\dot{V}$ and the parameters $p_0$ and $E$ across different fluid pressures $p_{\rm a}$. 
    At lower $p_{\rm a}$, the influence of changes in $p_0$ and $E$ on $\dot{V}$ is more pronounced than at higher $p_{\rm a}$. 
    For example, at $p_{\rm a} = 0 \ {\rm MPa}$, the difference in $\dot{V}$ between the prediction using $E = 1.8 \ {\rm MPa}$ and those using $E^+$ and $E^-$ spans approximately one and two orders of magnitude, respectively. In contrast, at $p_{\rm a} = 0.414 \ {\rm MPa}$, the corresponding differences reduced to about $4$ times higher and $1/9$ lower, respectively. 
    Similar behavior is observed when varying $p_0$. This sensitivity is attributed to the fact that near the percolation threshold, which corresponds to low $p_{\rm a}$ with negligible lift-off, the leak rate $\dot{V}$ is highly sensitive to small changes in the size of the leak paths, which depend on both $p_0$ and $E$. In contrast, under conditions where large interfacial separations are caused by high $p_{\rm a}$, the effect of asperity deformation becomes less significant.

    \item The errors caused by scaling $E$ or $p_0$ highlight the importance of accurate FEM results when predicting leakage using Persson contact mechanics. In traditional FEM simulations, errors of 15\% are common and often considered acceptable. However, our results show that such errors can lead to variations in predicted leak rates by several orders of magnitude.

    \item Within the framework of linear elasticity, and assuming that $p_{\rm a}$ does not affect the contact pressure (i.e., lift-off effects are negligible), leakage becomes independent of $E$ when compression is fixed (i.e., the geometry of the stopper and barrel remains unchanged). This is because increasing $E$ raises the contact pressure, but simultaneously reduces the ability to conform to asperities in rubber. As a result, surface separation depends on the ratio $p/E$, making leakage independent of $E$ \cite{huon2022air}. This behavior is illustrated by the red diamonds and lines in Fig. \ref{leak_rate_varp}, where both $p_0$ and $E$ are increased by $15\%$. As seen, $\dot V$ remains unchanged until $p_{\rm a}$ becomes large enough to induce lift-off.

    \item To improve sealing performance, reducing $E$ and increasing $p_0$ yield similar effects across most $p_{\rm a}$ values, except at very high $p_{\rm a}$ (e.g., Fig. \ref{leak_rate_varall} at $p_{\rm a} = 0.414 \ {\rm MPa}$). Therefore, the most effective method for reducing leakage is to increase the contact pressure, for instance by tightening the seal.
\end{enumerate}

\section{Conclusion}

In this study, we employed Persson contact mechanics theory, in combination with measured surface roughness spectra, contact pressure distributions obtained from FEM simulations, and relevant material properties, to predict leakage in a rubber stopper–glass barrel system.

The surface roughness of the rubber was characterized using two engineering stylus instruments and AFM, providing reliable topography measurements over different length scales. These measurements were used to compute the two-dimensional PSD of the surface, which is a central input in Persson contact mechanics theory.

To account for the complex geometry of the stopper ribs and the asymmetric contact pressure distribution, FEM simulations were performed. These calculations also incorporated the effect of internal fluid pressure on the deformation of the stopper, which alters the contact pressure distribution at the interface. While FEM assumes full contact between surfaces, in reality, fluid can penetrate into non-contact areas and increase the average surface separation under pressure, a phenomenon known as lift-off. This effect was taken into account in MCM calculations, enabling an accurate prediction of leakage rates.

Experimental leakage tests were conducted under controlled conditions, including both siliconized and cleaned (nominally dry) systems. Good agreement was found between the theoretical predictions and experimental results for dry contacts. For siliconized systems, no leakage was observed, which was attributed to the presence of capillary bridges formed by silicone oil and the delay due to the squeeze-out of silicon oil.

Additionally, a sensitivity analysis was carried out to investigate the influence of the elastic modulus $E_{\rm eff}$ and contact pressure $p_0 (x)$ on leakage. Results demonstrated that both parameters strongly affect leakage rates, particularly near the percolation threshold. The analysis also revealed that small deviations in input parameters, such as those typically considered acceptable in FEM simulations, can lead to large variations in the predicted leakage rate. This highlights the importance of accurate input data when applying contact mechanics models for seal design and evaluation.

Overall, this study contributes to a better understanding of the interplay between surface roughness, material properties, and sealing performance, and offers a predictive tool to support the design and optimization of reliable sealing systems in pharmaceutical and engineering applications.

%\section{Acknowledgment}
\vspace{2cm}

{\bf Appendix A: Fluid Flow Conductivity}

\begin{figure}[!ht]
\includegraphics[width=0.8\columnwidth]{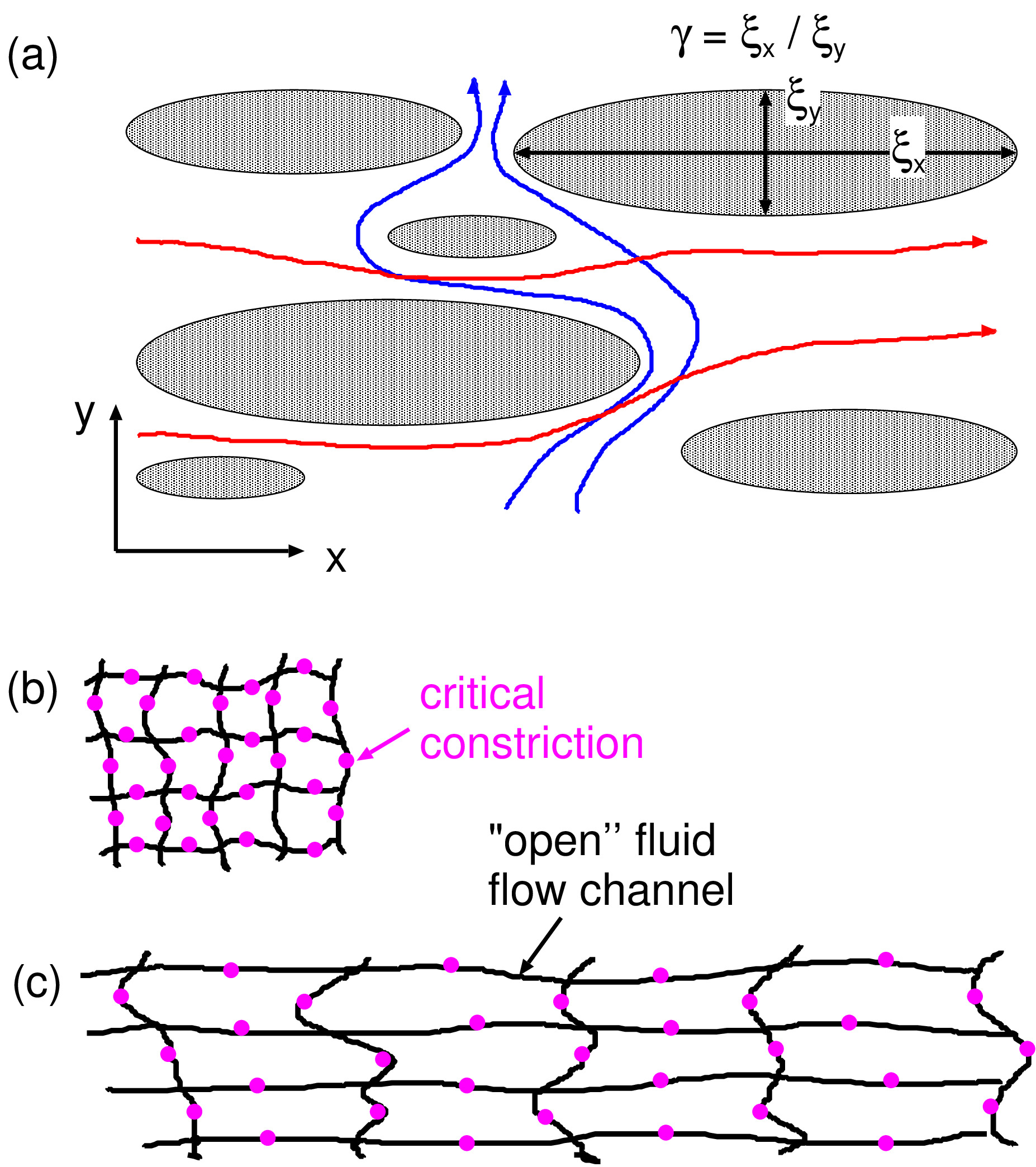}
\caption{\label{PERCOLATE.eps}
(a) Fluid flow in a contact with anisotropic roughness with Trip number $\gamma = 4$ (schematic).
The fluid flow conductivity in the $x$-direction is larger than that in the $y$-direction.
(b) Percolating (open) fluid flow channels (black lines) with narrow (critical) constrictions (pink dots), across which most
of the fluid pressure drop occurs.
(c) If the system in (b) is stretched by a factor $\lambda$ (here $\lambda = 4$) in the $x$-direction,
the number of fluid flow channels per unit area in the $x$-direction remains unchanged, but the concentration of
critical junctions along these channels decreases by a factor $1/\lambda$, which increases the
flow conductivity $\sigma_x$ by a factor $\lambda$. After stretching, the number of fluid flow channels per unit surface area
in the $y$-direction decreases by a factor $1/\lambda$, while the concentration of critical constrictions along these channels remains unchanged.
As a result, the flow conductivity $\sigma_y$ decreases by a factor $1/\lambda$. Therefore, for the stretched
system, $\sigma_x = \gamma^2 \sigma_y$.
}
\end{figure}

We consider the flow of a fluid (liquid or gas) at the interface between two elastic solids with anisotropic roughness characterized by the Tripp number $\gamma$, 
which is the ratio between the average size of the asperities (and valleys) in the $x$ and $y$ directions (see Fig. \ref{PERCOLATE.eps}). 

For this case, in a coordinate system where the flow conductivity tensor is diagonal, the effective medium theory predicts:
$${1\over \sigma_x} = \biggl \langle {1+(n-1)\gamma^* \over \sigma(u) +\gamma^* (n-1)\sigma_x } \biggr \rangle \eqno(A1)$$
$${1\over \sigma_y} = \biggl \langle {1+(n-1)(1/\gamma^*) \over \sigma(u) +(1/\gamma^*) (n-1)\sigma_y } \biggr \rangle \eqno(A2)$$
where the effective Tripp number $\gamma^* = (\sigma_y/\sigma_x)^{1/2} \gamma$. Here, $\langle .. \rangle$ denotes averaging over the probability distribution $P(u)$ of the interfacial separation, e.g.,
$$\langle f(u) \rangle = \int_0^\infty P(u) f(u)$$

In (A1) and (A2), the number $n$ appears, which in the Bruggeman effective medium theory corresponds to the dimension $D$ of the fluid flow problem. For the interfacial flow considered here, $D = 2$. However, using $n = 2$ results in percolation of the contact area when $A/A_0 = 0.5$, while experiments show that for randomly rough surfaces, percolation occurs at $A/A_0 \approx 0.42$. This discrepancy can be accounted for by using a dimension $n$ slightly less than 2 [see (A6) below, which gives $n \approx 1.75$]. We have found that using this value of $n$ yields leakage rates in almost perfect agreement with first-principles calculations of the fluid flow, which are feasible for small systems with roughness extending over approximately two decades in length scale \cite{dapp2012self}.

If $A/A_0$ denotes the normalized contact area, then we can write:
$$P(u)= {A\over A_0} \delta (u)+ P_{\rm c} (u) \eqno(A3)$$
where $P_{\rm c}(u)$ is the contribution from the surface region where $u > 0$. Substituting (A3) into (A1) and (A2), and using that $\sigma(0) = 0$, gives:
$${1\over \sigma_x} = {A_{\rm p}\over A_0} \left [{1+(n-1)\gamma^* \over \gamma^* (n-1)\sigma_x}\right ]+$$
$$\int_0^\infty du \  {1+(n-1)\gamma^* \over \sigma(u) +\gamma^* (n-1)\sigma_x } P(u) \eqno(A4)$$
$${1\over \sigma_y} = {A_{\rm p}\over A_0} \left [{1+(n-1)(1/\gamma^*) \over (1/\gamma^*) (n-1)\sigma_y } \right ]+$$
$$\int_0^\infty du \  {1+(n-1)(1/\gamma^*) \over \sigma(u) +(1/\gamma^*) (n-1)\sigma_y } P(u) \eqno(A5)$$

When the contact area percolates, no fluid can flow at the interface and the flow conductivity vanishes. Note that for a surface with random roughness, both $\sigma_x$ and $\sigma_y$ vanish when the contact area percolates. This is trivially true for a system with isotropic roughness, but also holds for a system with anisotropic roughness.

This can be understood in the limiting case where the anisotropic roughness is obtained by stretching in the $x$-direction a system with isotropic roughness. If the contact area percolates before the stretching, it must also percolate after the stretching. And if it does not percolate before stretching, it cannot percolate afterward.

From (A4) and (A5), as $\sigma_x \rightarrow 0$ and $\sigma_y \rightarrow 0$, we get:
$${A_{\rm p}\over A_0} \left [{1+(n-1)\gamma^* \over \gamma^* (n-1)}\right ]=1$$
$${A_{\rm p}\over A_0} \left [{1+(n-1)(1/\gamma^*) \over (1/\gamma^*) (n-1)}\right ]=1$$
It follows that $\gamma^* = 1$ and
$${A_{\rm p}\over A_0} \left [{1+(n-1) \over n-1} \right ]=1$$
or
$$n = {1\over 1 - A_{\rm p}/A_0} \eqno(A6)$$

The condition $\gamma^* = 1$ implies that $\sigma_x = \gamma^2 \sigma_y$, so that close to the percolation threshold, we expect the fluid current in the $x$-direction to be $\gamma^2$ times higher than in the $y$-direction. The physical reason why the flow current is larger in the $x$-direction than in the $y$-direction is illustrated in 
Fig. \ref{PERCOLATE.eps}(b,c).

The flow conductivity can be obtained by numerical integration and iteration from (A1) and (A2). However, for large interfacial separation $\bar u$, this is not convenient, and $\sigma_x$ and $\sigma_y$ for large $\bar u$ can instead be determined analytically. The result depends on $\sigma(u)$. For liquids, $\sigma(u) \sim u^3$ [see (5)], and for this case, the asymptotic large-$u$ form of the flow conductivity was derived in Ref. \cite{Prodanov}. 

For ballistic gas flow, $\sigma(u) \sim u^2$ [see the second term in (6)], and the asymptotic large-$u$ form of the flow conductivity can be obtained as follows. For ballistic gas flow:
$$\sigma(u) = {\bar v u^2 \over 2 k_{\rm B}T}$$
which we write as $\alpha u^2$ for simplicity. If $\bar u = \langle u \rangle$ is the ensemble-averaged surface separation (i.e., the surface separation after averaging over the roughness), we write $u = \bar u + u_1$ where $\langle u_1 \rangle = 0$. For large surface separation, there is no contact between the surfaces, and the surface roughness is not elastically deformed; hence, we expect $\langle u_1^2 \rangle$ to equal the mean square roughness $h_{\rm rms}^2$ of the undeformed surface. Defining $\bar \sigma = \alpha \bar u^2$, we get:
$$\sigma = \alpha \left (\bar u + u_1\right )^2 = \bar \sigma + 2 \alpha \bar u u_1 + \alpha u_1^2 = \bar \sigma \left (1+2\epsilon +\epsilon^2 \right).$$
Here, $\epsilon = u_1/\bar u$ is small when $\bar u$ is large, with $\langle \epsilon \rangle = 0$ and $\langle \epsilon^2 \rangle = h_{\rm rms}^2/\bar u^2$.

Writing $\sigma_x/\bar \sigma = y$, then (A1) becomes:
$${1\over y} = \biggl \langle {1+(n-1)\gamma^* \over 1+2\epsilon +\epsilon^2 +\gamma^* (n-1)y } \biggr \rangle$$

We compute $y$ to second order in $\epsilon$ (the first-order term vanishes). Writing $\mu = 1/[1+\gamma^* (n-1)y]$, we get:
$${1\over y} = [1+(n-1)\gamma^*] \mu \biggl \langle {1 \over 1+2\epsilon \mu +\epsilon^2 \mu} \biggr \rangle$$

Expanding to second order in $\epsilon$ and using $\langle \epsilon \rangle = 0$:
$${1\over y} = [1+(n-1)\gamma^*] \mu \left [1- \langle \epsilon^2 \rangle \left (4\mu^2 - \mu \right ) \right ]$$
or
$${1\over y} = \left [{1+(n-1)\gamma^* \over 1+(n-1)\gamma^* y} \right ] \left [ 1- \langle \epsilon^2 \rangle \left (4\mu^2 - \mu \right )\right ] \eqno(A7)$$

This equation shows that to zeroth order in $\epsilon$, $y = 1$ or $\sigma_x = \bar \sigma$, as expected. Writing $y = 1 + a \langle \epsilon^2 \rangle$, we get to second order:
$${1+(n-1)\gamma^* \over 1+(n-1)\gamma^* y} = 1 - (n-1)\gamma^* a \mu_1 \langle \epsilon^2 \rangle$$
where $\mu_1 = 1/[1+(n-1)\gamma^*] = \mu$ at $y=1$. Using this in (A7):
$$1 - a \langle \epsilon^2 \rangle = \left [ 1 - (n-1)\gamma^* a \mu_1 \langle \epsilon^2 \rangle \right ] \left [ 1 - \langle \epsilon^2 \rangle (4\mu_1^2 - \mu_1) \right ]$$
or
$$a \left [ -1 + (n-1)\gamma^* \mu_1 \right ] = 4\mu_1^2 - \mu_1$$
so
$$a = 1 - 4\mu_1 = {(n-1)\gamma^* - 3 \over 1 + (n-1)\gamma^*}$$
Hence,
$$\sigma_x = \bar \sigma \left (1 + \left [ {(n-1)\gamma^* - 3 \over (n-1)\gamma^* + 1} \right ] \langle \epsilon^2 \rangle \right ) \eqno(A8)$$

A similar expression with $\gamma^*$ replaced by $1/\gamma^*$ results for $\sigma_y$. Since both $\sigma_x$ and $\sigma_y$ are equal to $\bar \sigma$ to zeroth order in $\epsilon$, we can replace $\gamma^* = (\sigma_y/\sigma_x)^{1/2} \gamma$ with $\gamma$ in (A8) and the corresponding equation for $\sigma_y$. Using this and replacing $\langle \epsilon^2 \rangle$ with $h_{\rm rms}^2/\bar u^2$ gives:
$$\sigma_x = \bar \sigma \left (1 + \left [ {(n-1)\gamma - 3 \over (n-1)\gamma + 1} \right ] {h_{\rm rms}^2 \over \bar u^2} \right ) \eqno(A9)$$  
$$\sigma_y = \bar \sigma \left (1 + \left [ {(n-1)/\gamma - 3 \over (n-1)/\gamma + 1} \right ] {h_{\rm rms}^2 \over \bar u^2} \right ) \eqno(A10)$$  

For the case of a liquid, $\sigma \sim u^3$ [see (5)], and a similar derivation gives the asymptotic (large $\bar u$) relations:
$$\sigma_x = \bar \sigma \left (1 + \left [ {3[(n-1)\gamma - 2] \over (n-1)\gamma + 1} \right ] {h_{\rm rms}^2 \over \bar u^2} \right ) \eqno(A11)$$  
$$\sigma_y = \bar \sigma \left (1 + \left [ {3[(n-1)/\gamma - 2] \over (n-1)/\gamma + 1} \right ] {h_{\rm rms}^2 \over \bar u^2} \right ) \eqno(A12)$$  

The expression for $\sigma_x$ was derived in Appendix A of Ref. \cite{Prodanov} for the case $n = 2$ (a misprint in the derivation was overlooked during proofreading, but the final result is correctly stated).

%\bibliographystyle{apsrev4-2}
%\bibliography{Leakage.bib}

\end{document}